\documentclass{PoS}

\usepackage{amsmath}
\usepackage{booktabs}
\usepackage{enumitem}
\usepackage[utf8]{inputenc}

\rightline{TTP18-023}

\title{Charm decays}

\ShortTitle{Charm decays}

\author{\speaker{Stefan de Boer}
\\
        Karlsruhe Institute of Technology\\
        E-mail: \email{stefan.boer@kit.edu}}

\abstract{
The importance of charm decays in the context of flavor is motivated.
A theory overview on recent developments in leptonic and semileptonic decays, hadronic two-body decays, and rare decays is presented.
Standard Model (SM) predictions as well as tests of the SM are pointed out, {\it e.g.}~how to discover direct CP violation in charm decays and how to search for physics beyond the SM.
Differences between rare charm and beauty decays are discussed.
}

\FullConference{The International Conference on B-Physics at Frontier Machines - BEAUTY2018\\
		6-11 May, 2018\\
		La Biodola, Elba Island, Italy}

\begin{document}

\section{Introduction}

Why talk about charm decays at the BEAUTY conference?
We try to shed light on the link between beauty and charm in this proceedings.
While $b$ decays are established to test the standard model (SM), charm decays are commonly considered to be blurred by non-calculable hadronic effects.
These two members of the flavor family are driven by different dynamics, hence the way to search for physics beyond the SM (BSM) may point into different directions.
In this sense, charm and beauty are complementary in BSM searches and both can learn from each other.
One may compare the current situation of rare charm decays with that of the $b$(ig) brother back twenty years, which resulted in the exciting times nowadays.
Furthermore, charm is unique to test flavor in the up-type sector and allows for insights into QCD from a different perspective.
Experimentally, the $b$ machines are also charm machines \cite{li,williams,mezzadri,dicanto}.
On the theoretical side one should be careful in adopting results from $b$ physics, {\it e.g.}~the $1/m_c$-counting is questionable and the short-long distance behavior is challenging.
Additionally, the available phase space in charm decays is smaller, hence, {\it e.g.}, decays into $\tau$ leptons are suppressed or even forbidden.

We present a theory overview on charm decays for a selection of recent developments in this field.
Section \ref{sec:semileptonic} is an overview of current results of leptonic and semileptonic decays and determinations of the corresponding SM parameters.
In section \ref{sec:twobodyhadronic} we present the latest global fit to branching ratio data of hadronic two-body decays and outline how to test the SM with the potential to discover direct CP violation in charm decays.
Rare decays, reviewed in section \ref{sec:rare}, allow to search for BSM physics.
Section \ref{sec:summary} closes with the summary.

\section{Leptonic and semileptonic decays}\label{sec:semileptonic}

Leptonic and semileptonic $D\to l\nu$ and $D\to Pl\nu$ decays, where $P$ denotes a pseudoscalar meson allow to determine several SM parameters.
The decay into a $l=\tau$ lepton is kinematically suppressed/forbidden.
The corresponding decay amplitudes read
\begin{align}
 \mathcal A(D\to l\nu)&\propto V_{cq}^*\langle 0|\bar q\gamma_\mu\gamma_5 c|D(p')\rangle=V_{cq}^*[i p_\mu'f_D]\,,\label{eq:Dtolnu}\\
 \mathcal A(D\to Pl\nu)&\propto V_{cq}^*\langle P(p)|q\gamma_\mu c|D(p')\rangle\nonumber\\
 &=V_{cq}^*\left[f_+(q^2)\left((p'+p)_\mu-\frac{m_D^2-m_P^2}{q^2}q_\mu\right)+f_0(q^2)\frac{m_D^2-m_P^2}{q^2}q_\mu\right]\,,\label{eq:DtoPlnu}
\end{align}
where $q^2=(p'-p)^2$, the CKM factors $V_{cq}$, the decay constant $f_D$ and the form factors $f_{+,0}(q^2)$.
The amplitude (\ref{eq:DtoPlnu}) implies a lepton-mass-suppression of $f_0$ in charm decays, hence $f_0$ is presently barely accessible in experiments.
The CKM factors are given as $V_{cd}=\lambda+\mathcal O(\lambda^4)$ and $V_{cs}=1-\tfrac{\lambda^2}2+\mathcal O(\lambda^4)$ with $\lambda\sim0.225$ in the Wolfenstein parametrization.
Singly-Cabibbo suppressed $|\Delta c|=|\Delta u|=1$ decays, that will be covered in the next sections, involve the products $V_{ud}V_{cd}^*\simeq-V_{us}V_{cs}^*$.
This approximate equality which reflects the GIM mechanism allows only small complex phases $\sim V_{ub}V_{cb}^*$.

As is evident from (\ref{eq:Dtolnu}) and (\ref{eq:DtoPlnu}), experiments extract products of CKM factors with decay constants and form factors from leptonic and semileptonic decays, respectively.
The individual quantities can then be obtained by either assuming CKM unitarity and the other CKM factors to be known or from a combination with, {\it e.g.}, lattice QCD (LQCD) calculations.
In practice, the CKM factors are most precisely known from fits to several observables employing CKM unitarity, hence we do not consider their extraction in the following.

The interplay of experiments and LQCD calculations in the determination of the decay constants is summarized in Table \ref{tab:fD}.
\begin{table}
 \centering
 \begin{tabular}{ccc}
  \toprule
  & $f_D\,[\text{MeV}]$ & $f_{D_s}\,[\text{MeV}]$ \\
  \midrule
  HFLAV 2016 & $203.7(4.9)$ & $257.1(4.6)$ \\
  FLAG 2016 & $212.15(1.45)$ & $248.83(1.27)$ \\
  \bottomrule
  \end{tabular}
 \caption{Decay constants from experiments, assuming CKM unitarity, and $N_f=2+1+1$ LQCD calculations, summarized by HFLAV \cite{Amhis:2016xyh} and FLAG \cite{Aoki:2016frl}, respectively.}
 \label{tab:fD}
\end{table}
The averaged results from experiments and LQCD calculations are compatible at $2\sigma$ with competing but smaller uncertainties from the LQCD calculations.
In addition, recent results from $N_f=2+1$ and $N_f=2+1+1$ calculations with individual uncertainties similar to the 2016 averaged ones are provided by RBC/UKQCD \cite{Boyle:2017jwu} and Fermilab Lattice/MILC \cite{Bazavov:2017lyh}, respectively.
While lattice calculations need to include QED effects non-perturbatively in the future, also more precise experimental measurements are needed for further comparisons.
Decay constants are also determined by QCD sum rule calculations, however, with larger uncertainties than the results collected in Table \ref{tab:fD}, {\it e.g.}~$f_D=(208\pm10)\,\text{MeV}$ and $f_{D_s}=(240\pm10)\,\text{MeV}$ \cite{Wang:2015mxa}.

\begin{figure}
 \begin{center}
  \includegraphics[trim=2cm 8cm 4cm 4cm,clip,width=0.4\textwidth]{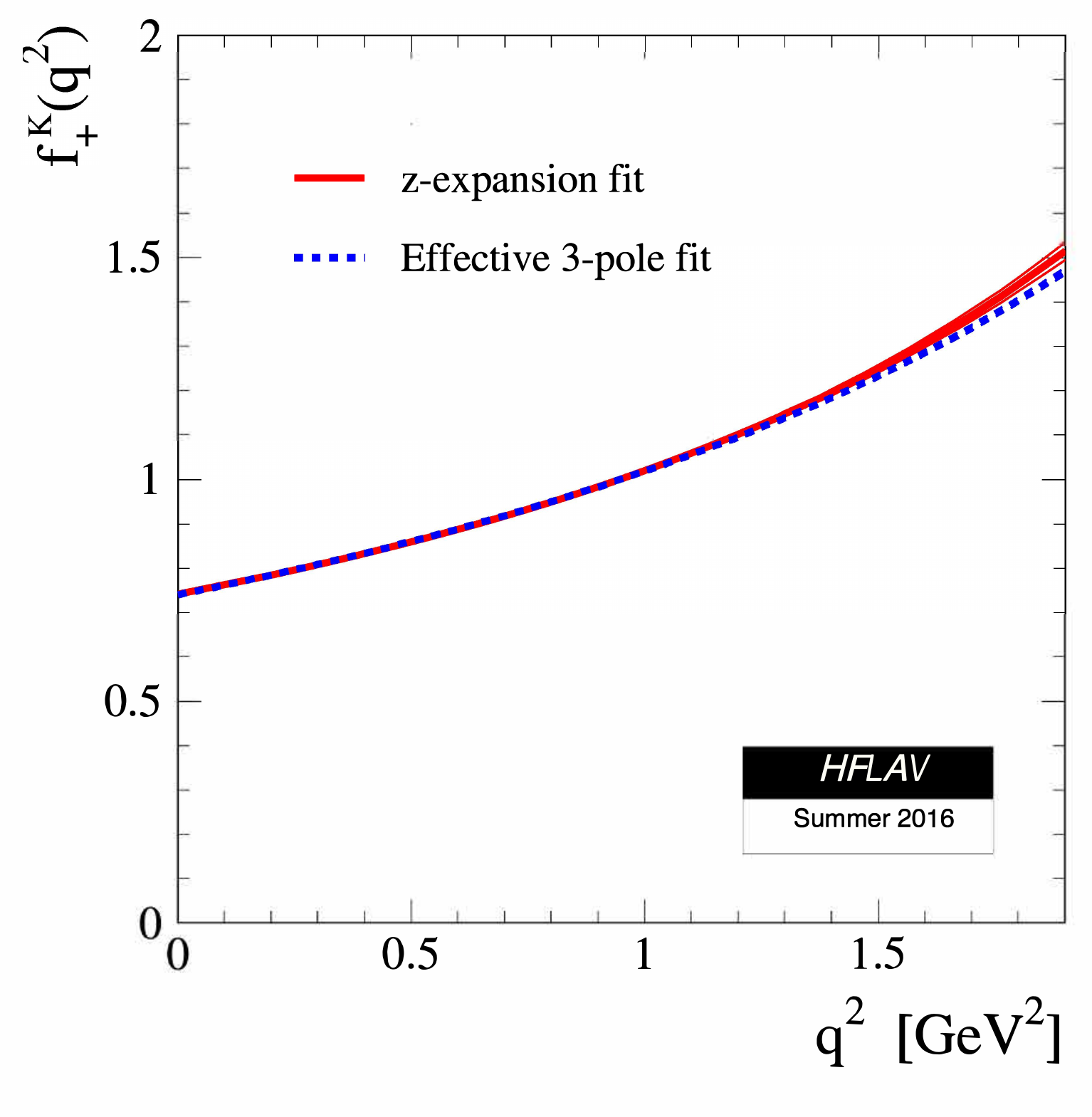}\quad
  \includegraphics[trim=2cm 8cm 4cm 4cm,clip,width=0.4\textwidth]{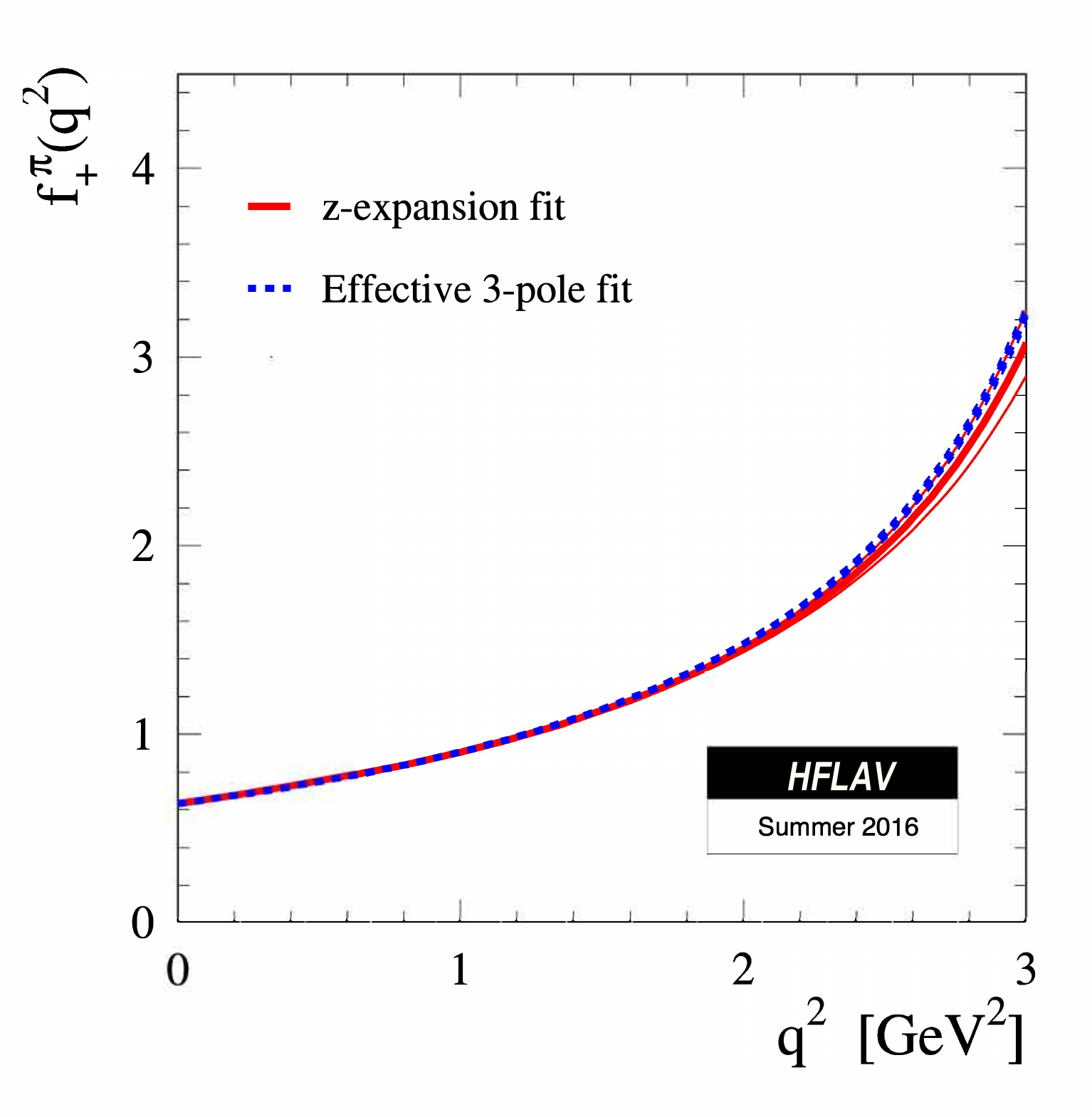}
 \end{center}
 \caption{Form factor $f_+$ for $D\to(K,\pi)$, averaged by HFLAV.
 Figure taken from \cite{Amhis:2016xyh}.}
 \label{fig:fplus}
\end{figure}

In Figure \ref{fig:fplus}, we show the experimental average of the form factor $f_+$ for $D\to(K,\pi)$.
While experiments precisely extract form factors at low $q^2$, form factors at high $q^2$ are accurately calculated on the lattice.
In fact, the experimental results are in agreement with, {\it e.g.}, the recent $N_f=2+1+1$ calculation provided by ETMC \cite{Lubicz:2017syv}.
The interplay of experiments and LQCD calculations allows to determine form factors at the percent level precision and is also necessary in view of $f_0$ that is presently only accessible on the lattice.

Finally, we note that the form factors for $\Lambda_c\to N$, where $N$ is a nucleon, are available from a recent LQCD calculation \cite{Meinel:2017ggx}.

\section{Hadronic two-body decays}\label{sec:twobodyhadronic}

Hadronic two-body decays, $D\to P_1P_2$ with $P_{1,2}\in\{\pi,K\}$ have been studied extensively, see {\it e.g.}~\cite{Brod:2012ud,Hiller:2012xm,Khodjamirian:2017zdu}, including first attempts on the lattice \cite{Hansen:2012tf}.
Most of the branching ratios, their correlations and also CP-asymmetries are experimentally measured \cite{Amhis:2016xyh,Patrignani:2016xqp}.
Theoretically, amplitudes for different decays can be related by the $SU(3)_F$-symmetry, allowing to test the SM.
The amplitude for, {\it e.g.}, a singly-Cabibbo suppressed decay is written as
\begin{align}
 \mathcal A=\lambda_{sd}\,A_{sd}-\frac{\lambda_b}2\,A_b
\end{align}
with $\lambda_{sd}=\tfrac{\lambda_s-\lambda_d}2$ and $\lambda_q=V_{cq}^*V_{uq}$.
Recall that, while $\lambda_b$ can be neglected for branching ratios, it is the only source of CP violation in the SM.
In the following, we summarize the global fit of branching ratios and the CP-asymmetry predictions given in the series of works \cite{Muller:2015lua,Muller:2015rna,Nierste:2015zra,Nierste:2017cua}.

The framework consists of topological amplitudes with diagrammatic $SU(3)_F$-breaking and color-counting input.
A global fit to branching ratio data reveals that the $SU(3)_F$-limit is excluded by more than $5\sigma$, while $30\%$ $SU(3)_F$-breaking in the decay amplitudes is sufficient to describe the data \cite{Muller:2015lua}.
Predictions on individual branching ratios are obtained as $\mathcal B(D_s^+\to K_LK^+)=0.012_{-0.002}^{+0.006}$ at $3\sigma$ CL and $\mathcal B(D^0\to K_L\pi^0)<\mathcal B(D^0\to K_S\pi^0)$ with a significance of more than $4\sigma$ \cite{Muller:2015lua}.

For the direct CP violation, the corresponding asymmetry is written as
\begin{align}
 a_{CP}^\text{dir}=\mathrm{Im}\frac{\lambda_b}{\lambda_{sd}}\,\mathrm{Im}\frac{A_b}{A_{sd}}.
\end{align}
Here, $\mathrm{Im}\frac{\lambda_b}{\lambda_{sd}}\simeq-6\times 10^{-4}$ and $|A_{sd}|$ can be taken from the branching ratio fit.
However, CP-asymmetries require additional combinations of amplitudes that are not provided by a branching ratio fit.
Nevertheless, these combinations can be eliminated by sum rules that correlate different CP-asymmetries.
To address charm CP violation, two strategies are given at hand: Firstly, the SM can be falsified with sum rules, or clean predictions.
Secondly, direct CP violation can be discovered in charm, with large SM predictions being favored.
An example for the first strategy is $A_{CP}(D^+\to\pi^+\pi^0)\simeq0$ from isospin sum rules, {\it e.g.}~\cite{Grossman:2012eb}, which is compatible with the recent Belle measurement $a_{CP}^\text{dir}=+0.0231\pm0.0124\pm0.0023$ \cite{Babu:2017bjn}.

The discovery of direct CP violation is shown to be possible in $D^0\to K_SK_S$ decays with $|a_{CP}^\text{dir}|\le1.1.\%$ \cite{Nierste:2015zra} that follows from sizable tree level exchange, and since $A_{sd}=0$ in the $SU(3)_F$-limit, while $A_b\ne0$.
Experimentally, $A_{CP}=-0.0002\pm0.0154$ is measured by Belle \cite{Dash:2017heu} with uncertainties being dominated by statistics as well as very recently LHCb measured $A_{CP}=0.042\pm0.034\pm0.010$ \cite{Aaij:2018jud}.
Direct CP violation can also be discovered in $D\to K_S{K^*}^0$ decays, $|a_{CP}^\text{dir}|\le0.3\%$ \cite{Nierste:2017cua}.
While the features of the $D^0\to K_SK_S$ decay also apply here, the decay $D\to K_S{K^*}^0$ is additionally experimentally favored, since charged tracks from prompt $K_S{K^*}^0$ decay can be identified, a Dalitz plot analysis may reveal regions with large strong phases to maximize the asymmetry, and no flavor tagging is required.
No measurements for this asymmetry is presently available, though a first study is reported in \cite{Aaij:2015lsa}.

\section{Rare decays}\label{sec:rare}

Rare charm decays are very different from the analogous $b$ decays, nevertheless $|\Delta c|=|\Delta u|=1$ decays allow to test BSM physics in addition to and independent of their $b$ counterpart.
We first review the anatomy of the SM, the background in BSM searches.
The perturbative SM contribution is described by the Lagrangian $\mathcal L_\text{eff}^\text{weak}\sim\sum_iC_iP_i$ with Wilson coefficients $C_i$ and the corresponding operators, {\it e.g.},
\begin{align}
 &P_{2{\color{blue}(1)}}\sim(\bar u_L\gamma_\mu{\color{blue}(T^a)}q_L)(\overline q_L\gamma^\mu{\color{blue}(T^a)}c_L)\,,\nonumber\\
 &P_{\color{black}7}^{\color{blue}(\prime)}\sim(\bar u_{L{\color{blue}(R)}}\sigma^{\mu\nu}c_{R{\color{blue}(L)}})F_{\mu\nu}\,,\nonumber\\
 &P_{9{\color{blue}(10)}}\sim(\bar u_L\gamma_\mu{\color{blue}}c_L)(\bar\ell\gamma^\mu{\color{blue}(\gamma_5)}\ell)\,,
\end{align}
where the notation and the full set of operators can be found in, {\it e.g.}, \cite{deBoer:2015boa}.
One characteristic of charm flavor-changing-neutral-current transitions is the two-step matching of the effective theory at the scales $m_W$ and $m_b$.
Another one is related to the absence of heavy down-type quarks in loops, hence only two Wilson coefficients ($C_{1,2}$) are nonzero at the scale $m_W$ and $C_{10}\simeq0$ holds at the charm scale that is broken only by, {\it e.g.}, electromagnetic effects.
The effective Wilson coefficients, that include perturbative contributions which multiply the same matrix elements as the Wilson coefficients, are known at the same order as in $b$ physics \cite{Greub:1996wn,Fajfer:2002gp,deBoer:2016dcg,deBoer:2017way}.
The phenomenologically largest contributions to $C_{7,9}^\text{eff}$ result from two-loop QCD diagrams with insertions of $P_{1,2}$ \cite{deBoer:2017way}, since the leading order contributions are generically suppressed by the GIM-mechanism due to closely degenerated light quarks in the loops.
Note that the two-loop calculation \cite{deBoer:2017way} is valid for arbitrary momentum transfer and also for $b$ decays.

\begin{figure}
 \begin{center}
  \includegraphics[width=0.8\textwidth]{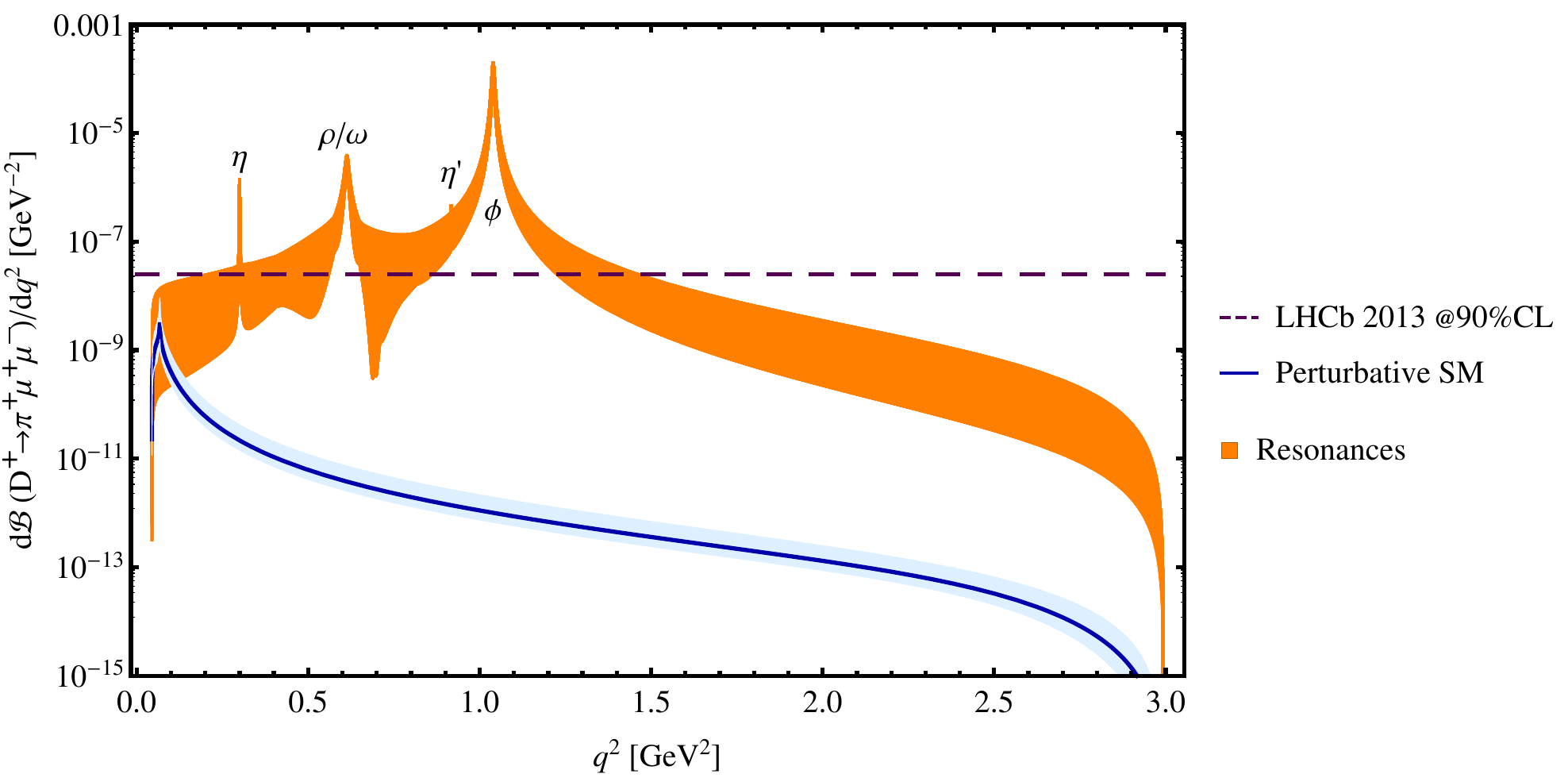}
 \end{center}
 \caption{Differential branching ratio with respect to $q^2$ for $D^+\to\pi^+\mu^+\mu^-$.
 Figure is an update of \cite{deBoer:2015boa}.}
 \label{fig:dBQDppiR2mu}
\end{figure}

\begin{figure}
 \begin{center}
  \includegraphics[width=0.8\textwidth]{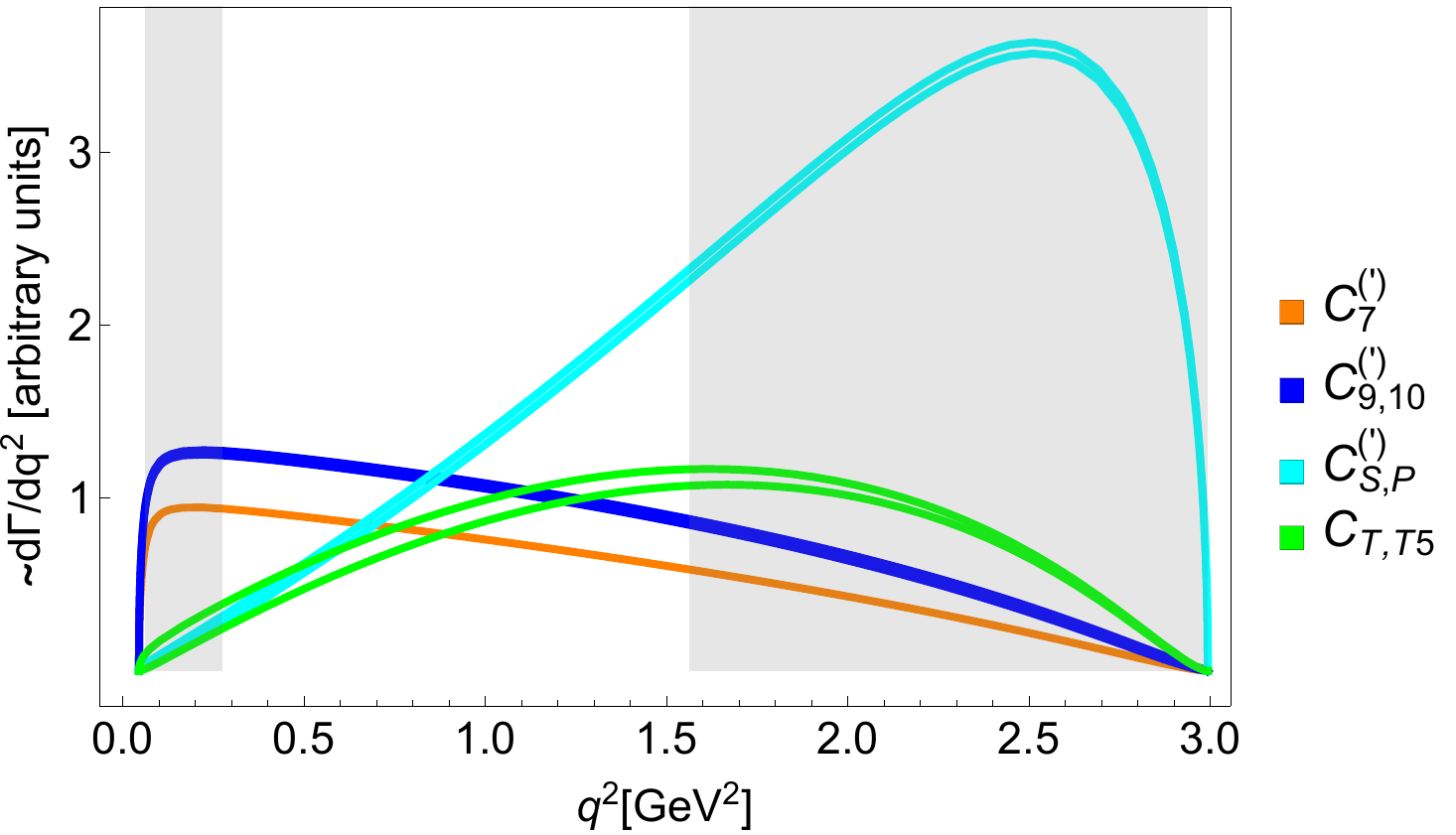}
 \end{center}
 \caption{Comparison of short-distance spectrum sensitivities to different Wilson coefficient in $D^+\to\pi^+\mu^+\mu^-$.
 Figure taken from \cite{Fajfer:2015mia}.}
 \label{fig:sens}
\end{figure}

Switching from partons to hadrons, intermediate resonances evade the GIM-mechanism and dominate the branching ratios, though their contribution is uncertain, {\it e.g.}~\cite{Feldmann:2017izn}.
The different contributions to the branching ratios are summarized for $D^+\to\pi^+\mu^+\mu^-$ in Figure \ref{fig:dBQDppiR2mu}.
Charm decays may not look viable to search for BSM physics due to the large and uncertain SM background.
On the other side, several unique SM features arise/persist, {\it e.g.}, so-called ``resonance-catalyzed'' \cite{Fajfer:2012nr} observables as well as BSM sensitive observables due to the small SM weak phases and symmetries of QCD and QED.
They allow to search for (heavy) BSM physics in different ways:
\begin{enumerate}[label=(\alph*)]
 \item Windows in branching ratios, {\it e.g.}~the high $q^2$ region in Figure \ref{fig:dBQDppiR2mu}.
 \item Null test based on (approximate) SM symmetries.
 \item SM contribution extracted from SM-dominated modes as input for $SU(3)_F$-related, BSM sensitive modes.
\end{enumerate}
Before illustrating these points, we emphasize to look into different decays and observables to probe the SM and sort BSM models.
Figure \ref{fig:sens} shows the sensitivity of the differential branching fraction for $D^+\to\pi^+\mu^+\mu^-$ to different Wilson coefficients.

Another example for (a) is the branching ratio of $D^0\to\mu^+\mu^-$.
The present experimental upper limit $\mathcal B_\text{exp}<6.2\times10^{-9}$ by LHCb {\cite{Aaij:2013cza} sets the strongest constraints on the difference of (pseudo)scalar Wilson coefficients.
While the SM branching ratio is commonly estimated orders of magnitude below $\mathcal B_\text{exp}$ \cite{Burdman:2001tf,Fajfer:2001ad,Paul:2010pq}, a BSM-induced branching ratio close to the experimental limit is allowed within, {\it e.g.}, two Higgs doublet and leptoquark models \cite{deBoer:2015boa,Fajfer:2015mia,Burdman:2001tf,Paul:2010pq,Golowich:2009ii,Paul:2012ab,Wang:2014uiz}.
The decay into electrons is, on the other hand, helicity suppressed and diluted by misidentification from $\mathcal O(\alpha m_D^2/m_e^2)$ enhanced $D^0\to e^+e^-\gamma$ decays with soft photons \cite{Fajfer:2002gp}.
The helicity suppression is lifted for ${D^0}^*\to e^+e^-$ decays, however ${D^0}^*$ decays strongly/electromagnetically.
Note that the decay ${D^0}^*\to\mu^+\mu^-$ is also diluted by misidentification from ${D^0}^*\to\pi^+\pi^-$ decays.
As explored in \cite{Khodjamirian:2015dda}, $e^+e^-\to{D^0}^*$ decays evade these complications.
The SM branching ratio of $e^+e^-\to{D^0}^*$ is predicted to be $\mathcal B_\text{SM}\sim10^{-18}$, while in $Z'$ models $\mathcal B_{Z'}<2.5\times10^{-11}$ \cite{Khodjamirian:2015dda}.
A measurement would probe (axial)vector Wilson coefficients, not accessible in $D^0\to ll$ decays.

Generically, for $c\to u\ell\ell^{(\prime)}$ induced decays, {\it e.g.}~$D\to P\ell\ell$ and $D\to PP\ell\ell$, SM CP-asymmetries are small, $A_{CP}^\text{SM}\sim\tfrac{\mathrm{Im}\,\lambda_b}{\lambda_s}\sim10^{-3}$.
Further SM null test are given in terms of angular observables, {\it e.g.}~the dilepton forward-backward asymmetry, which involve $C_{10}$, (pseudo)scalar and tensor Wilson coefficients and are suppressed in the SM.
Additionally, the SM can be testes with decays into different leptons.
Lepton-universality, probed by the ratio of muons/electrons, holds up to percentage corrections in the SM \cite{Fajfer:2015mia}.
Experimentally, decays into electrons and muons are measured in different experiments, {\it e.g.}~\cite{Aaij:2017iyr,Ablikim:2018gro} employed different cuts for $D\to PPll$ decays.
Lepton-flavor-violating decays, also from quarkonia, are absent in the SM \cite{deBoer:2015boa,Burdman:2001tf,Hazard:2016fnc,Hazard:2017udp}, hence are further null tests of the SM.
Decays into neutrinos which vanish in the SM and also probe dark matter \cite{deBoer:2015boa,Burdman:2001tf,Paul:2012ab,Badin:2010uh} are bounded by Belle \cite{Lai:2016uvj}.
Predictions for the CP-asymmetry and the forward-backward asymmetry within BSM models are compiled in Table \ref{tab:ACP_AFB_BSM}, ranging from ``SM-like'' to ``within reach of the next measurement''.
\begin{table}
 \centering
 \begin{tabular}{ccc}
  \toprule
  model  &  $A_{CP}$  &  $A_{FB}$  \\
  \midrule
  Leptoquark models  &  $\gtrsim A_{CP}^\text{SM}$  &  $\lesssim8\times10^{-1}$  \\[0.1em]
  Little Higgs model  &  $\lesssim\mathcal O(10^{-3})$  &  $\lesssim\mathcal O(5\times10^{-3})$  \\[0.1em]
  Minimal SUSY SM  &  $\lesssim\mathcal O(10^{-3})$  &  $\lesssim\mathcal O(10^{-1})$  \\[0.1em]
%  Two Higgs doublet model  &  --  &  --  \\[0.1em]
%  Unparticle  &  --  &  --  \\[0.1em]
  Up vector-like quark singlet  &  --  &  $\lesssim10^{-3}$  \\[0.1em]
  Warped extra dimension  &  $\lesssim\mathcal O(10^{-2})$  &  $\lesssim\mathcal O(5\times10^{-2})$  \\[0.1em]
%  Weak vector triplet  &  --  &  --  \\[0.1em]
  Z' boson  &  --  &  $\sim0$  \\[0.1em]
  \midrule
  SM  &  $<\mathcal O(10^{-3})$  &  $\sim0$ \\
  \bottomrule
  \end{tabular}
 \caption{CP asymmetry and forward-backward asymmetry within BSM models.
 Table is a compilation of the results given in \cite{deBoer:2015boa,Fajfer:2015mia,Burdman:2001tf,Paul:2012ab,Wang:2014uiz,Fajfer:1998rz,Fajfer:2001sa,Fajfer:2005ke,Fajfer:2006yc,Fajfer:2007dy,Fajfer:2008tm,Paul:2011ar,Bigi:2011em,Delaunay:2012cz,Cappiello:2012vg,Guo:2017edd,Sahoo:2017lzi}.}
 \label{tab:ACP_AFB_BSM}
\end{table}

The following example of a ``resonance-catalyzed'' \cite{Fajfer:2012nr} CP-asymmetry comprises several of the just discussed qualities.
Consider a scalar leptoquark (with quantum numbers (3,3,-1/3)), supplement a flavor pattern (inspired by $b$ decays \cite{Varzielas:2015iva}) and respect constraints from Kaon decays (due to $SU(2)_L$ couplings).
This recipe results in Figure \ref{fig:ACPphirhoDppi2mu_phi_0004}, illustrating that the CP-asymmetry around resonances probes the Wilson coefficient $\mathrm{Im}\,C_9^\text{BSM}$, independent of strong phases.
\begin{figure}
 \begin{center}
  \includegraphics[width=0.8\textwidth]{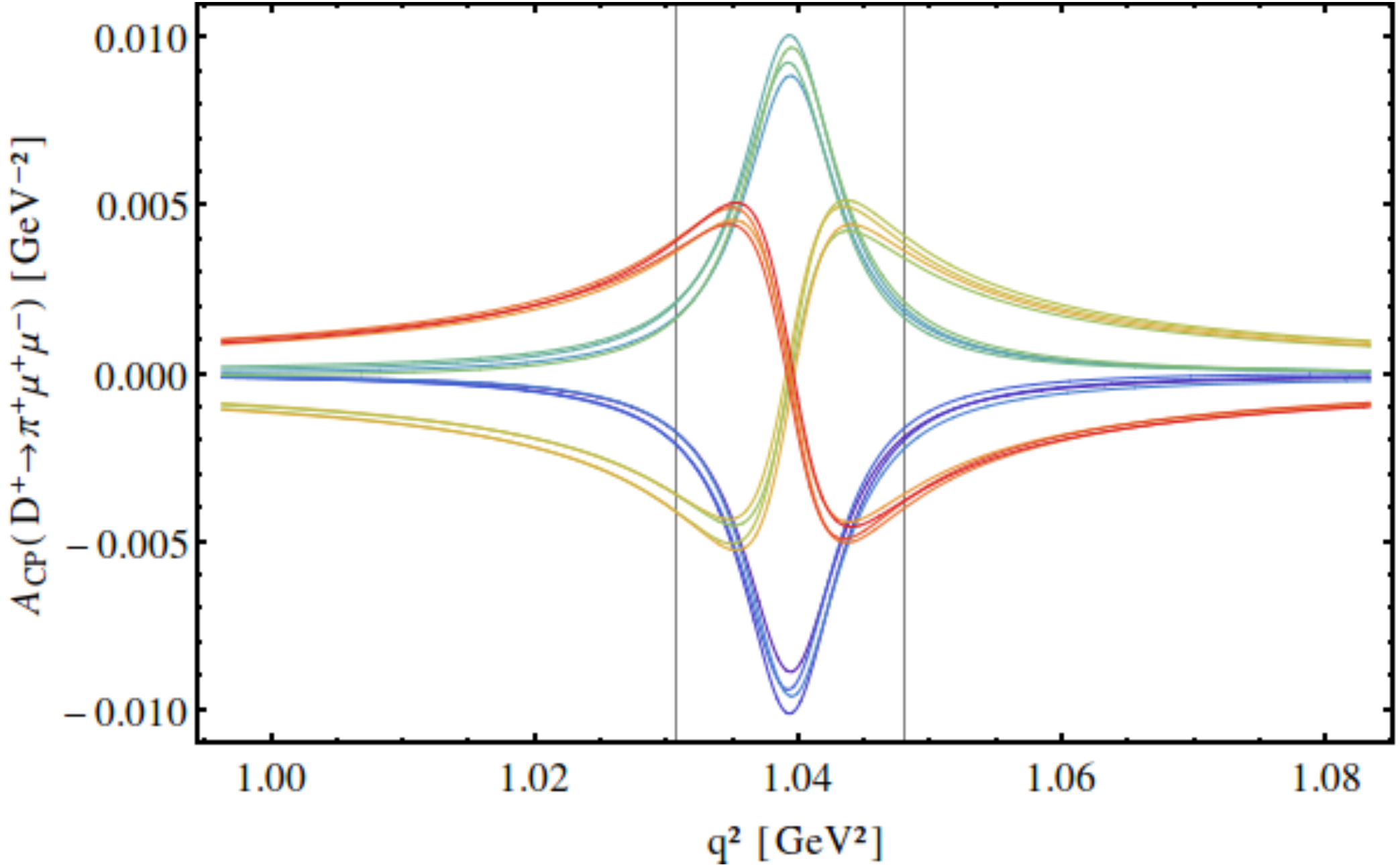}
 \end{center}
 \caption{Resonance-catalyzed CP-asymmetry for a scalar leptoquark model, around the $\phi$ resonance, normalized to the shown bin and for the strong phases (${\color{yellow}\pi/2},{\color{green}\pi},{\color{blue}0},{\color{red}3/2\pi}$).
 Figure taken from \cite{deBoer:2015boa}.}
 \label{fig:ACPphirhoDppi2mu_phi_0004}
\end{figure}

Finally, we discuss radiative $c\to u\gamma$ induced decays.
Again, branching ratios are dominated by uncertain long-distance effects \cite{Fajfer:2002gp,Burdman:1995te,Khodjamirian:1995uc,Fajfer:1997bh,Fajfer:1998dv,Prelovsek:2000xy,Dimou:2012un,deBoer:2017que,Biswas:2017eyn,Dias:2017nwd}.
On the other hand, the decay $B_c\to B_u^*\gamma$ and the ratio $\Gamma(D^0\to\rho^0\gamma)/\Gamma(D^0\to\omega\gamma)$ are sensitive to BSM physics \cite{Fajfer:1999dq,Fajfer:2000zx}.
The SM prediction for CP asymmetries is, again, $A_{CP}^\text{SM}<\mathcal O(10^{-3})$, while in BSM models $A_{CP}^\text{BSM}\lesssim10\%$ \cite{deBoer:2017que,Isidori:2012yx,Lyon:2012fk}.
Here, the first experimental measurement of $A_{CP}^\text{exp}(D^0\to\rho^0\gamma)=0.056\pm0.152$, where the statistical uncertainty dominates, is obtained by Belle \cite{Abdesselam:2016yvr}.

A feature of radiative decays is an observable photon polarization, the ratio of right/left-handed currents ($C_7'/C_7$).
The photon polarization can be probed in the following ways.
\begin{itemize}
 \item Time-dependent analysis: Relate the SM-dominated decay $\bar D^0\to\bar{K^*}^0\gamma$ to the decays $\bar D^0\to(\rho^0/\omega,\phi)\gamma$ using data and $SU(3)_F$-symmetry and extract the BSM contribution \cite{Lyon:2012fk,deBoer:2018zhz}.
 \item The up-down asymmetry of the decay $D\to\bar K_1(\to\bar K\pi\pi)\gamma$ is independent of strong phases between $C_7$ and $C_7'$ and heavier resonances are phase space suppressed, but $D$-tagging is required \cite{deBoer:2018zhz}.
 \item The photon forward-backward asymmetry in the decay $\Lambda_c\to p\gamma$ can be measured at future colliders \cite{deBoer:2017que}.
\end{itemize}

\section{Summary}\label{sec:summary}

Charm decays allow for a wide analysis of the SM and beyond.
On the one hand, SM parameters, namely decay constants and form factors, are precisely known from leptonic and semileptonic decays.
Recent experiments and LQCD computations challenge each other with competing uncertainties.
On the other hand, hadronic two-body decays allow for global fits to branching ratio data and to test the $SU(3)_F$-symmetry.
Direct CP violation of charm decays in the SM can be discover in $D^0\to K_SK_S$ and $D\to K_S{K^*}^0$ decays with SM CP asymmetries $\lesssim1\%$.

Rare charm decays allow to uniquely probe the SM and BSM physics with different decays and observables -- despite branching ratios being dominated by long-distance effects.
Examples are CP-asymmetries and angular observables in semileptonic decays, decays into different leptons as SM null test and connections between different radiative decays.
While BSM physics may link flavor sectors, one should not forget that charm decays may also improve our understanding of QCD and check theoretical frameworks.
Several experiments and theoretical works are ongoing, {\it e.g.}~on $D\to PPll$ decays \cite{deBoer:2018buv,Aaij:2018fpa}.

\section*{Acknowledgements}

I thank the organizers for the invitation and the wonderful conference.
I am grateful to Ulrich Nierste, Stefan Schacht and Michele Della Morte for valuable discussion.
This work has been supported by the BMBF under contract no.~05H15VKKB1.
I acknowledge the DAAD for financial support to attend the conference.

\end{document}